\documentclass[aps,prl,twocolumn,groupedaddress,showpacs]{revtex4-1}

\usepackage{array}
\usepackage{amsmath,amssymb,amstext}
\usepackage{graphicx}
\usepackage{multirow}
\usepackage{subfigure}
\def\rvec{{\bf r}}
\def\kvec{{\bf k}}

\def\hvec{{\bf h}}

\begin{document}

\title{Dynamic structure function of a cold Fermi gas at unitarity}

\author{G. E. Astrakharchik$^{\ddagger}$, J.~Boronat$^{\ddagger}$,
  E.~Krotscheck$^{+\dagger}$ and T.~Lichtenegger$^{\dagger}$}
\affiliation{$^\ddagger$Departament de F\'\i sica i Enginyeria Nuclear, Campus
Nord B4-B5, Universitat Polit\`ecnica de Catalunya, E-08034 Barcelona, Spain}
\affiliation{$^+$Department of Physics, University at Buffalo, SUNY
Buffalo NY 14260}
\affiliation{$^\dagger$Institut f\"ur Theoretische Physik, Johannes
Kepler Universit\"at, A 4040 Linz, Austria}

\date{\today}

\begin{abstract}
  We present a theoretical study of the dynamic structure function of
  a resonantly interacting two-component Fermi gas at
  zero temperature. Our approach is based on dynamic many-body theory
  able to describe excitations in strongly correlated Fermi systems.
  The fixed-node diffusion Monte Carlo method is used to produce the
  ground-state correlation functions which are used as an input for
  the excitation theory. Our approach reproduces recent Bragg scattering
  data in both the density and the spin channel. In the BCS regime,
  the response is close to that of the ideal Fermi gas. On the BEC
  side, the Bose peak associated with the formation of
  dimers dominates the density channel of the dynamic response. When the
  fraction of dimers is large our theory departs from the experimental
  data, mainly in the spin channel.

\end{abstract}


\maketitle

An impressive advance in realizing and controlling ultracold Fermi
gases has permitted to study physical phenomena whose appearance was
previously only a matter of speculation~\cite{review}. One of the
major advances has been the first clean observation of a crossover
between two limits of Fermi matter, the Bose-Einstein condensate
(BEC) regime composed by Fermi molecules and the
Bardeen-Cooper-Schrieffer (BCS) superfluid gas where Cooper pairs are
formed~\cite{zwierlein,partridge,shin,navon}.

In the dilute limit, interactions between atoms can be described by a
single parameter, namely the $s$-wave scattering length $a$. The
system evolves from a BCS regime with no bound state ($a\,<\,0$) to a
molecular one (BEC) with a two-body bound state ($a\,>\,0$), crossing a
singular point where $|a| \rightarrow \infty$ corresponding to a
Fano-Feshbach resonance~\cite{astra,carlson,gandolfi}.  This special
point is referred to as the unitary limit and is expected to show a
number of universal properties. Indeed, in this regime the only
relevant length and energy scales are the inverse Fermi momentum
$k_F^{-1}$ and the Fermi energy $E_{\rm F}$, respectively.

Nowadays, the unitary Fermi gas is routinely realized in experiments.
An acccurate theoretical description of the unitary limit requires the
use of advanced methods of many-body theory that can deal with
strongly interacting systems.  In this situation, quantum Monte Carlo
methods have proven to be very useful for the calculation of the
finite- and zero-temperature equation of
state~\cite{astra,carlson,gandolfi,drut0} which is found to be in good
overall agreement with experimental data~\cite{salomon,mit}.  Also, it
was possible to predict the effect of the unbalanced
population~\cite{pilati} and different masses~\cite{grigori}.  More
recently, static responses such as the spin susceptibility and spin
diffusivity~\cite{sanner,sommer} have been measured and also studied
theoretically using quantum Monte Carlo (QMC) methods~\cite{drut} and
a diagrammatic approach~\cite{pieri}.

Recently, Hoinka \textit{et al.}~\cite{Hoinka2012,Hoinka2013} have
measured the dynamic spin and density responses of a Fermi gas at
unitarity and in the BEC regime. In order to measure both responses,
they used two-photon Bragg scattering, acting on $^6$Li at very low
temperature ($\sim 0.06 E_{\rm F}$ at unitarity), and a proper choice of the
laser detuning.  The density response at unitarity is significantly
different from that of an ideal Fermi gas, showing a clear peak which
corresponds to the formation of dimers composed of spin-up and
spin-down particles. This signature of dimer formation is not visible
in the spin channel, nevertheless our results are quite
different from the ideal Fermi gas response. This provides evidence for
the importance of correlations in the unitary gas despite its
diluteness.

Some of the features shown by the dynamic response function at
unitarity were previously determined in theoretical work: at $T=0$
using a dynamic self-consistent mean-field approach based on BCS
theory~\cite{combescot}, at $T>0$ using virial
expansions~\cite{hu}.  However, the dynamic properties along the
BCS-BEC crossover are not completely understood due to the need of a
full many-body theory and the impossibility to use QMC methods
directly. The goal of our study is to pursue further a theory for the
dynamic structure function, combining the virtues of Monte
Carlo methods and modern diagrammatic many-body theory.

In the present work, we study the density and spin responses of the
two-component Fermi gas at unitarity and in characteristic points on
both the BEC and BCS sides of the crossover.  We use a fully
microscopic approach utilizing correlated basis function (CBF) theory.
The input ground-state structure functions are obtained from diffusion
Monte Carlo simulations within the fixed-node (FN-DMC)
approximation~\cite{astra}. In its most advanced
form~\cite{kro_theory}, the dynamic many-body theory (DMBT) used here
is the fermion version of the CBF-Brillouin-Wigner perturbation theory
previously developed for bosons~\cite{campbell}. The power of DMBT for
fermions has been recently demonstrated by the prediction of a stable
roton excitation in two-dimensional $^3$He, which has independently
been confirmed by inelastic neutron scattering
experiments~\cite{godfrin}.

Inputs for the calculation of the dynamic response are the static
structure factors, $S_{\uparrow \uparrow}(k)$ and $S_{\uparrow
  \downarrow}(k)$, and the fraction of dimers along the BCS-BEC
crossover.  We obtain a complete description of the density and spin
dynamic responses in the momentum-energy plane. In the unitary limit,
both theoretical responses are in satisfactory agreement with the
recent experimental Bragg scattering measurements~\cite{Hoinka2012}. 

In weakly interacting systems, time-dependent Hartree-Fock (TDHF)
theory~\cite{Thouless1972} is a well-established method for capturing
the dynamics.  One assumes that the system is subjected to the
perturbing Hamiltonian
\begin{equation}
 \delta \hat{H}_{\text{ext}}(t)=
\int d^3r \hat{\rho}^{(\rho/\sigma)}(\rvec) h_{\text{ext}}(\rvec,t) \,;
\end{equation}
the superscripts $\rho$ and $\sigma$ stand for density and spin
excitations, respectively. $\hat{\rho}^{(\rho/\sigma)}(\rvec)$ is the
(spin-)density operator, and $h_{\text{ext}}(\rvec,t)$ is a weak, local
external field. The wave function resulting from such a
perturbation is assumed to be of the form
\begin{equation}
 |\psi\rangle=e^{1/2\sum_{ph} c_{ph}(t)a^{\dagger}_pa_h}|\phi_0\rangle\,,
\end{equation}
where $|\phi_0\rangle$ is the ground state of a non-interacting system
with the same density. The particle (``$p$'') and hole (``$h$'')
labels run over spatial quantum numbers and spin degrees of freedom.

The amplitudes $c_{ph}(t)$ are Fourier decomposed
\begin{equation}
 c_{ph}(t)=c^{(+)}_{ph}(\omega)e^{-\imath(\omega+\imath\eta/\hbar)t}
+c^{(-)*}_{ph}(\omega)e^{\imath(\omega+\imath\eta/\hbar)t}\,.
\end{equation}
Equations of motion are then derived from a least action
principle~\cite{KermanKoonin1976,KramerSaraceno1980}\,. They are given
by
\begin{widetext}
\begin{align}
 \big(\hbar\omega+\imath\eta-\epsilon_{ph}\big)c^{(+)}_{ph}(\omega)-
\sum_{p'h'}V_{ph;p'h'}c^{(+)}_{p'h'}(\omega)-
\sum_{p'h'}V_{pp'hh';0}c^{(-)}_{p'h'}(\omega)
=&2\int d^3r\rho^{(\rho/\sigma)}_{ph;0}(\rvec) h_{\text{ext}}(\rvec,\omega)\\
 \big(-\hbar\omega-\imath\eta-\epsilon_{ph}\big)c^{(-)}_{ph}(\omega)
-\sum_{p'h'}V_{p'h';ph}c^{(-)}_{p'h'}(\omega)-\sum_{p'h'}V_{0;pp'hh'}c^{(+)}_{p'h'}(\omega)
=&2\int d^3r\rho^{(\rho/\sigma)}_{0;ph}(\rvec) h_{\text{ext}}(\rvec,\omega),
\end{align}
\end{widetext}
where $ h_{\text{ext}}(\rvec,\omega)$ is the Fourier component
of the external field.

From the solutions $c^{(\pm)}_{ph}(\omega)$ we can calculate the
(spin-) density response function
\begin{equation}
 \chi^{(\rho/\sigma)}(\rvec,\rvec',\omega)\equiv\frac{1}{\rho(\rvec)\rho(\rvec')}
\frac{\delta \left\langle \psi\right|\hat{\rho}^{(\rho/\sigma)}
\left|\psi\right\rangle(\rvec,\omega)}
{\delta h_{\text{ext}}(\rvec',\omega)} \ ,
\end{equation}
and the (spin-)density dynamic structure function
\begin{equation}
 S^{(\rho/\sigma)}(\rvec,\rvec',\omega)
\equiv-\frac{1}{\pi}\text{Im}\chi^{(\rho/\sigma)}(\rvec,\rvec',\omega)
\end{equation}
in coordinate space. In a translationally invariant geometry,
their momentum space representations,
$\chi^{(\rho/\sigma)}(q,\omega)$ and $S^{(\rho/\sigma)}(q,\omega)$,
are functions of the momentum transfer $q$.

Key ingredients of the equations of motion are the matrix elements
of the interparticle interaction
\begin{align}
 V_{ph;p'h'}&=
\left\langle ph'\right|V_{\text{d}}\left|hp'\right\rangle
-\left\langle ph'\right|V_{\text{ex}}\left|p'h\right\rangle\\
 V_{pp'hh';0}&=
\left\langle pp'\right|V_{\text{d}}\left|hh'\right\rangle
-\left\langle pp'\right|V_{\text{ex}}\left|h'h\right\rangle
\end{align}
and the single-particle spectrum
$\epsilon_{ph}=\epsilon_p-\epsilon_h$. Generally, the potentials
$V_{\text{d}}$ and $V_{\text{ex}}$ are effective interactions in the
direct and exchange channel, respectively.  In conventional TDHF
theory, they are the same and just the bare interaction between the
particles. The $\epsilon_p$, $\epsilon_h$ are the Hartree-Fock
single-particle energies. Using the CBF method, the description can be
extended to strongly interacting
systems~\cite{ChenClarkSandler1982,Krotscheck1982}. The basic outcome
of the rather intricate diagrammatic analysis is that one arrives at
the same equations of motion. CBF theory provides a method
for computing effective interactions from the bare interaction.
These interactions are normally different in the
direct and the exchange channels.

If one neglects exchange effects, one can construct $V_{\text{d}}$ from
the static structure function by demanding that $V_{\text{d}}$
reproduces $S^{(\rho)}(q)$ through the $\omega^0$ sum rule
\begin{equation}
 S^{(\rho)}(q)=\int_0^{\infty} d(\hbar\omega)S^{(\rho)}(q,\omega) \ .
\end{equation}
A useful analytic relationship which is accurate within one to two percent
can be derived using a collective approximation for the Lindhard
function,
\begin{align}
 V_{\text{d}}(q)=\frac{\hbar^2q^2}{4m}\Big(\frac{1}{S^{(\rho)}(q)^2}-
\frac{1}{S_{\text{F}}(q)^2}\Big) \ ,
\label{potdir}
\end{align}
where $S_{\text{F}}(q)$ is the static structure factor of the free Fermi gas.

For spin-independent interactions, the spin-density fluctuations
depend on $V_{\text{ex}}(q)$ but not on $V_{\text{d}}(q)$. Thus one
might be tempted to construct $V_{\text{ex}}(q)$ from
$S^{(\sigma)}(q)$ in a similar fashion as $V_{\text{d}}(q)$ is derived
from $S^{(\rho)}(q)$. Unfortunately, this procedure has numerical
difficulties because the resulting problem is very poorly
conditioned. We therefore utilize an expression obtained from the
Fermi-Hypernetted-Chain (FHNC) analysis which suggests, in its
simplest form~\cite{KroFHNC}, a local exchange interaction

\begin{equation}
 V_{\text{ex}}(q)=-\frac{\hbar^2q^2}{2m} \,
\frac{S^{(\rho)}(q)-S_{\text{F}}(q)}{S_{\text{F}}(q)^3} \,.
\label{potex}
\end{equation}
The same effective interaction defines the single-particle spectrum
\begin{equation}
 \epsilon_k = \frac{\hbar^2k^2}{2m}-
\frac{1}{\nu N}\sum_h V_{\text{ex}}(|\kvec-\hvec|)\,,
\end{equation}
where $\nu$ is the spin degeneracy and $N$ the number of particles.
A consistent treatment of the exchange potential and the single particle
spectrum is needed to satisfy the $\omega^1$ sum rule.

The calculation of the potentials $V_{\text{d}}(q)$~(\ref{potdir}) and
$V_{\text{ex}}(q)$~(\ref{potex}) requires knowledge of no more than
the density structure function of the interacting system,
$S^{(\rho)}(q)$. For that, we use ground-state results obtained using
the FN-DMC method~\cite{astra}, which has proven its accuracy in the
description of the BCS-BEC crossover.

\begin{figure}
\begin{center}
\includegraphics[width=0.6\linewidth,angle=-90]{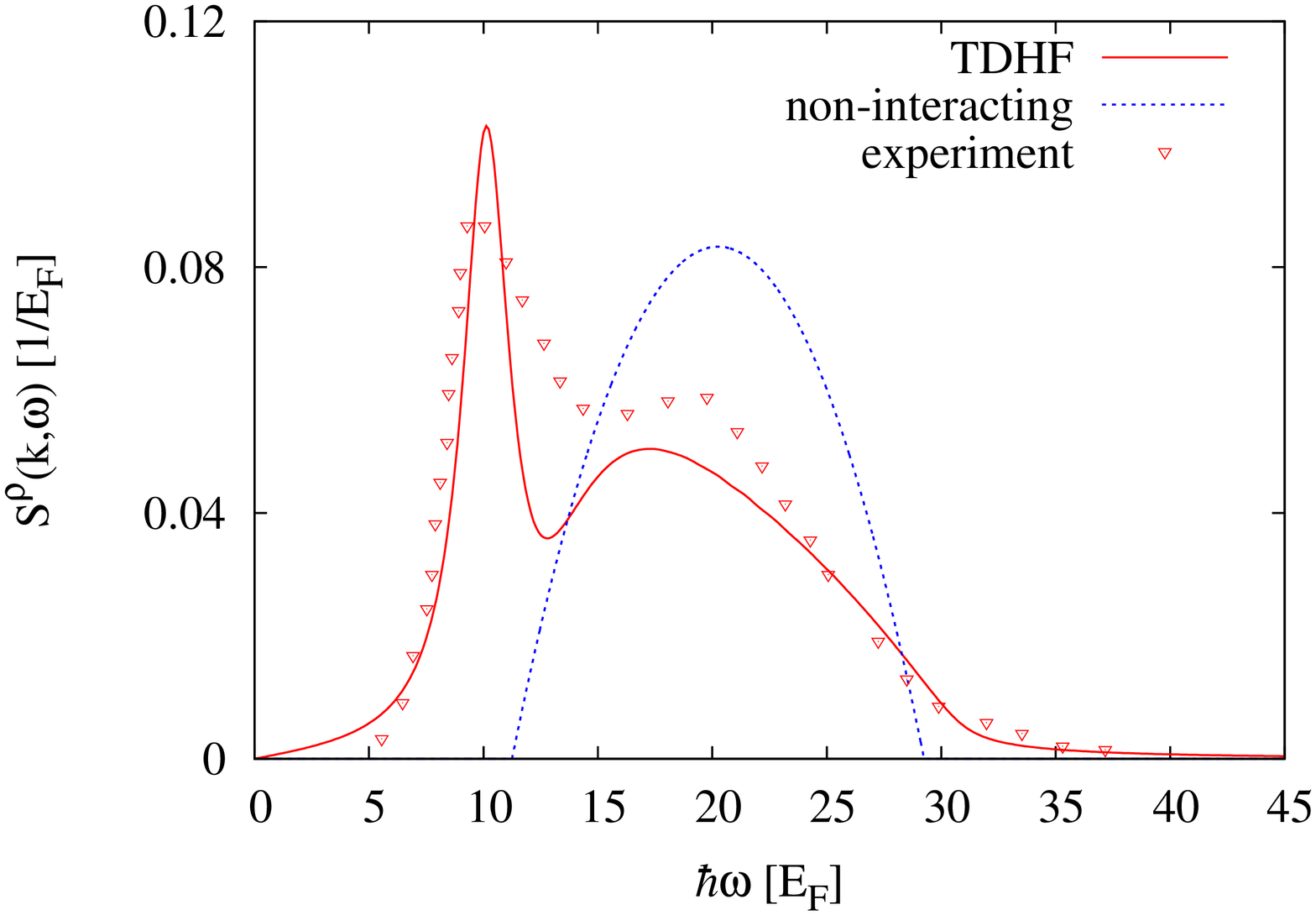}

\includegraphics[width=0.6\linewidth,angle=-90]{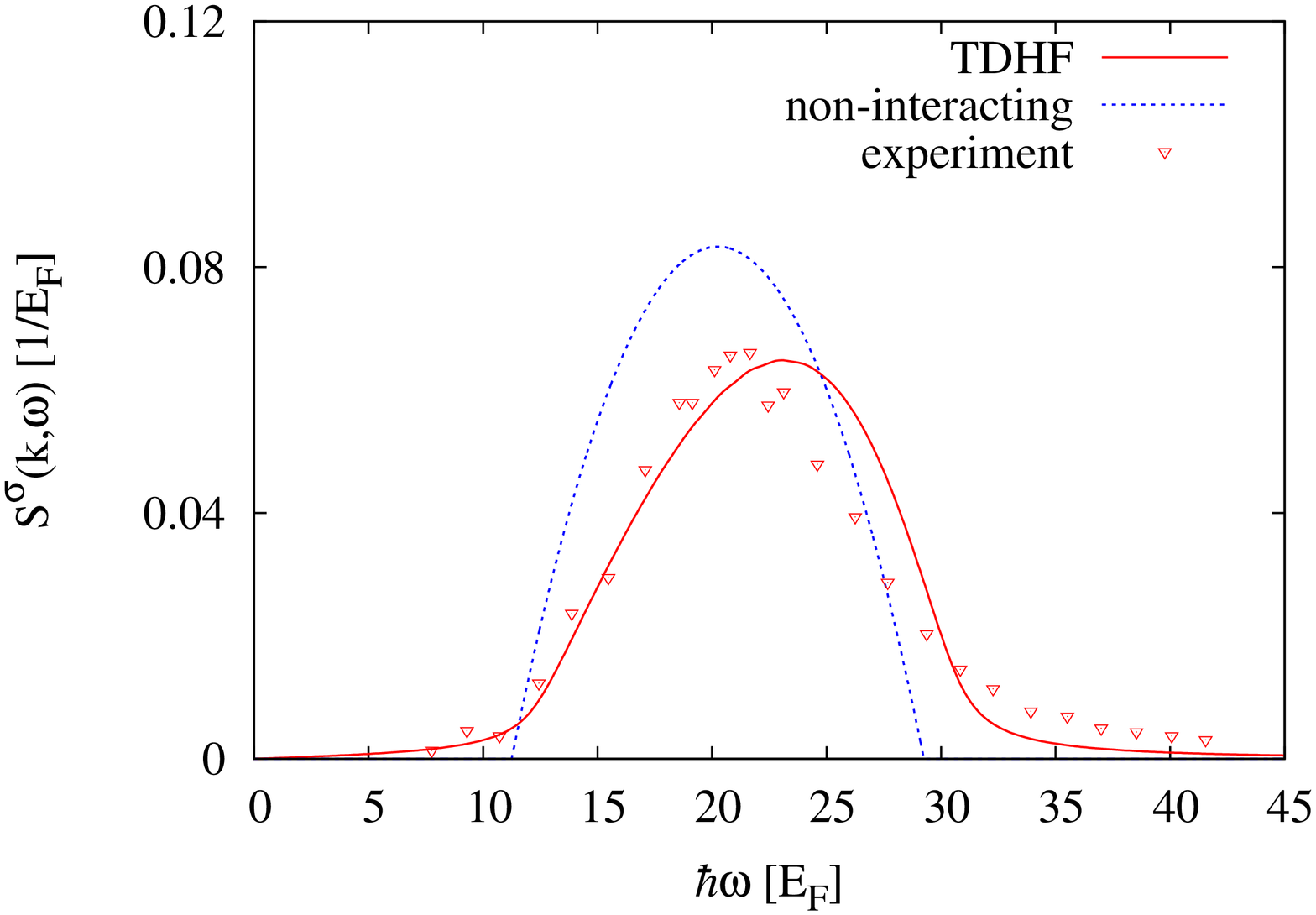}
\end{center}
\caption{(Color online) The dynamic structure function in the density
  (\textit{top}) and spin (\textit{bottom}) channels at unitarity
  ($1/(k_Fa)=0$) and at a momentum $q=4.5k_{\text{F}}$ (solid
  line). The points are the experimental data of
  Ref. \cite{Hoinka2012}. Dashed line stands for the non-interacting
  response. }
\label{fig1}
\end{figure}

We have calculated the dynamic structure function in the density and
spin channels for $1/(k_{\text{F}}a)=-1$ (BCS like),
$1/(k_{\text{F}}a)=0$ (unitarity) and $1/(k_{\text{F}}a)=1$ (BEC like)
including exchange terms and non-local, energy-dependent CBF
corrections that are not spelled out explicitly, see
Ref.~\cite{Krotscheck1982}.  Paying attention to the fact that,
with decreasing $1/(k_{\text{F}}a)$, the system favors the creation of
bound dimers, we consider a non-interacting mixture of dimers
(``bosons'' with $2m$) and atoms (fermions). The relative
concentrations by which we weight the two contributions to the dynamic
structure function in the density channel are obtained from our FN-DMC
ground-state calculations. In particular, the concentration of dimers
as a function of the scattering length is estimated assuming that all
pairs are in the zero-momentum state. According to this criterion, the
concentration of dimers is 100\% deep inside the BEC regime and
decreases to zero quickly after crossing the unitary limit where it
amounts $\sim$ 50\% ~\cite{astra2}. Our model cannot predict the width
of the ``Bose'' peak, hence it has been fitted to reproduce the width
of the experimental data subject to the constraint that its
contribution to the $\omega^0$ sum rule is constant.  The spin channel
dynamics on the other hand is not affected by the presence of dimers
because they do not contribute to spin-density fluctuations.

\begin{figure}
\begin{center}
\includegraphics[width=0.6\linewidth,angle=-90]{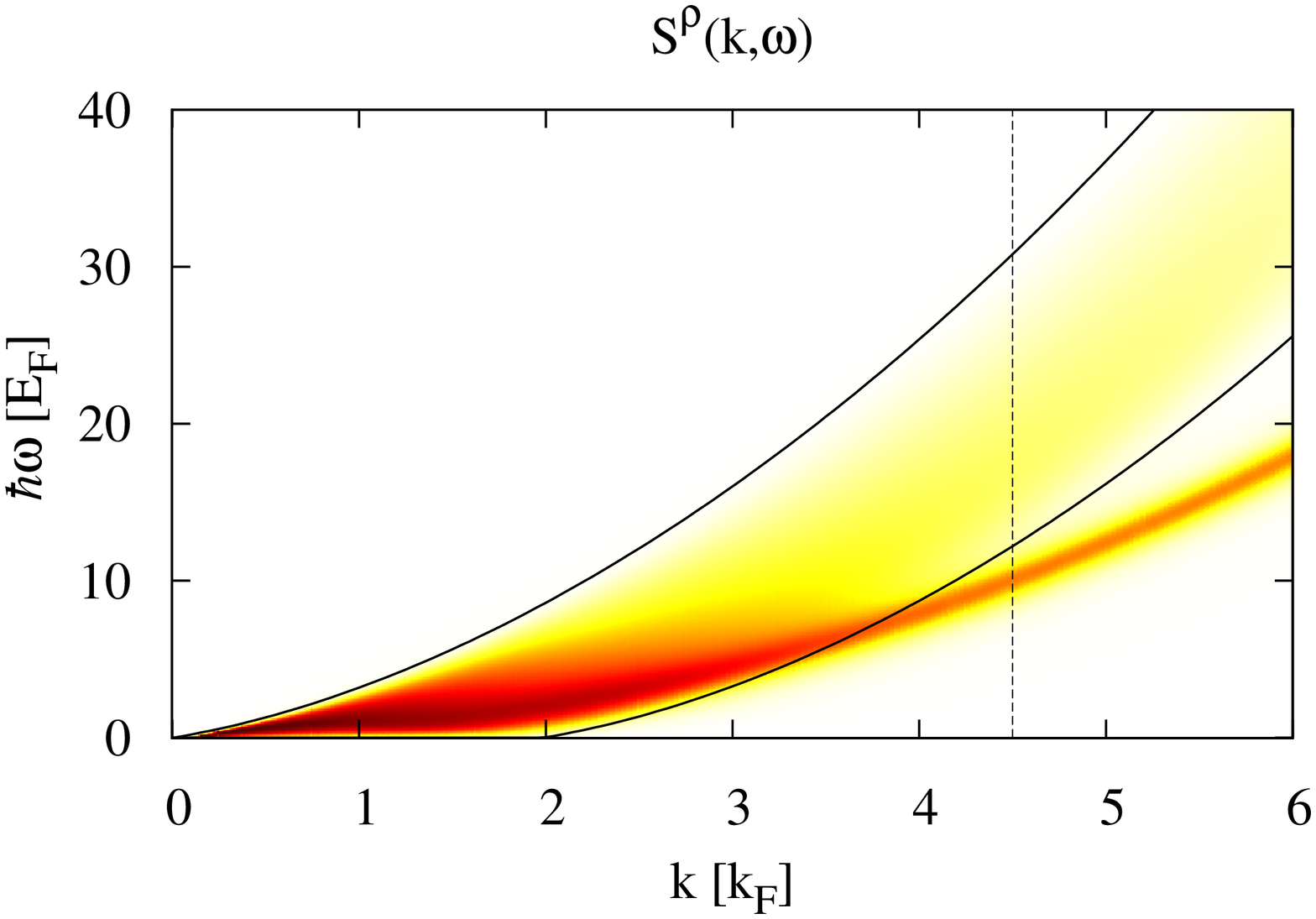}

\includegraphics[width=0.6\linewidth,angle=-90]{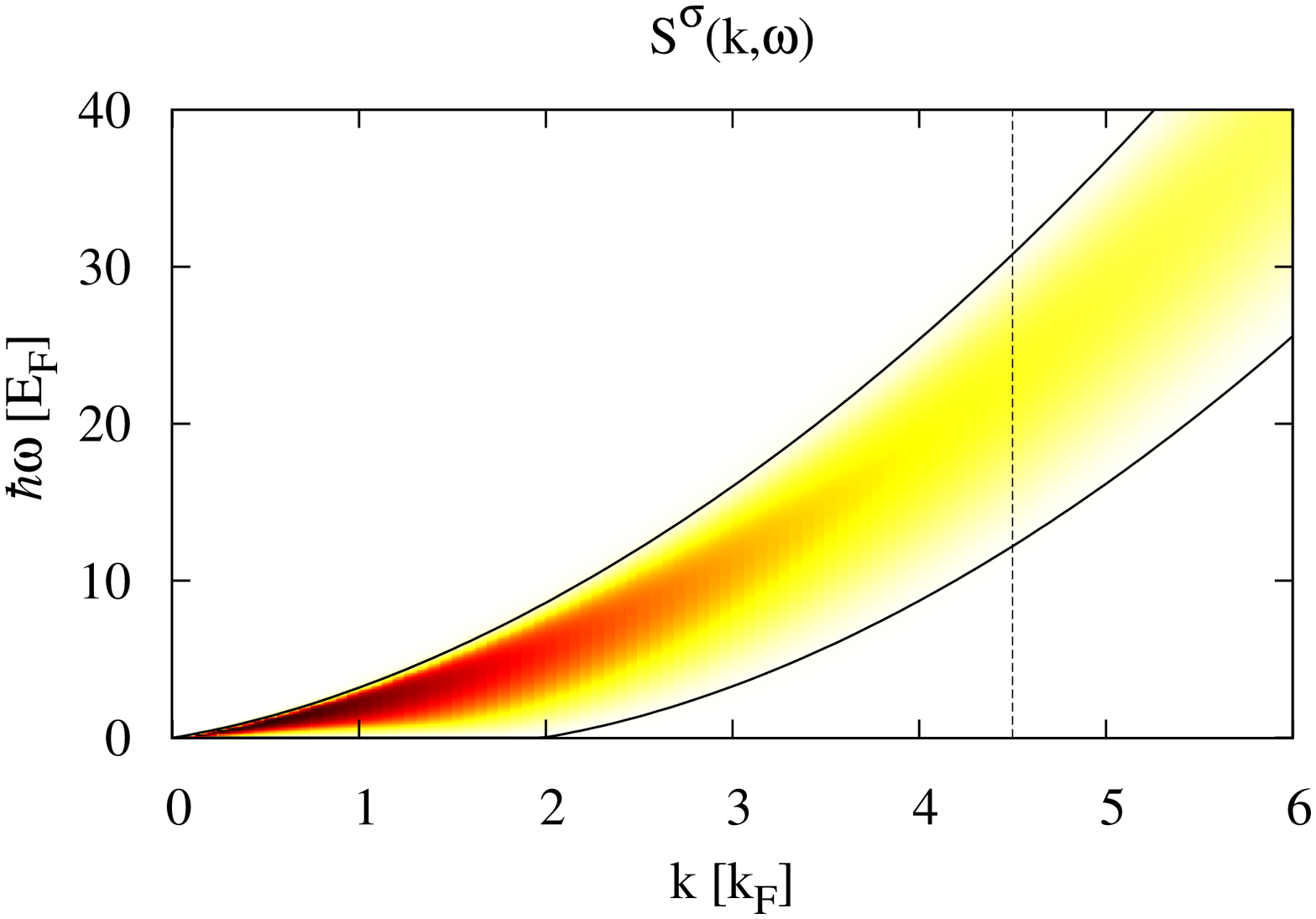}
\end{center}
\caption{(Color online) The dynamic structure function in the density
  (\textit{top}) and spin (\textit{bottom}) channels at unitarity
  ($1/(k_Fa)=0$) as a function of the momentum transfer $k$ and energy
  transfer $\hbar \omega$ in units of the Fermi energy $E_{\rm F}$.
  The solid lines show the limits of the particle-hole band. The
  vertical dashed line corresponds to the momentum transfer at which
  the experiment of Ref. \cite{Hoinka2012} was performed.}
\label{fig2}
\end{figure}

\begin{figure}
\begin{center}
\includegraphics[width=0.6\linewidth,angle=-90]{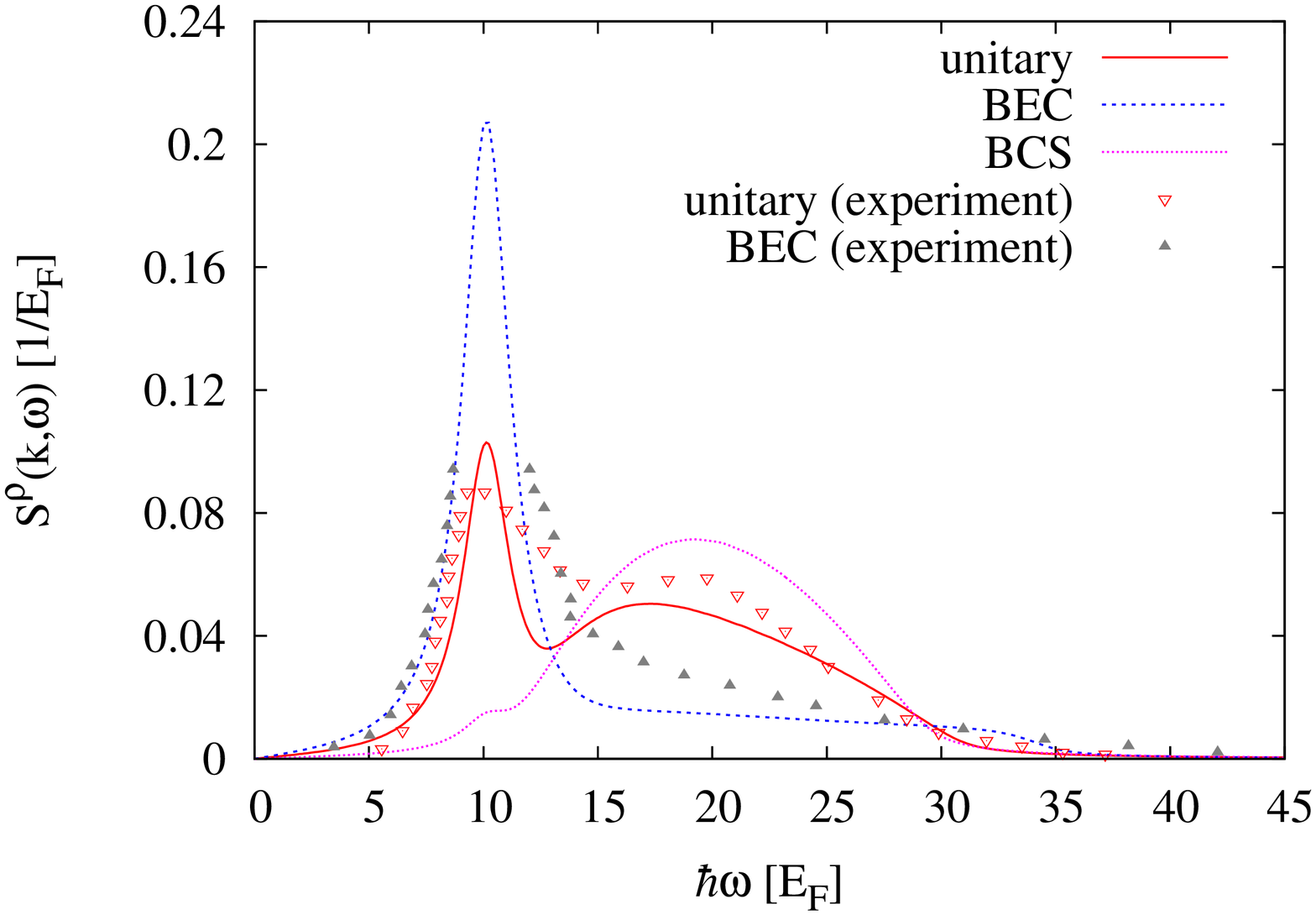}

\includegraphics[width=0.6\linewidth,angle=-90]{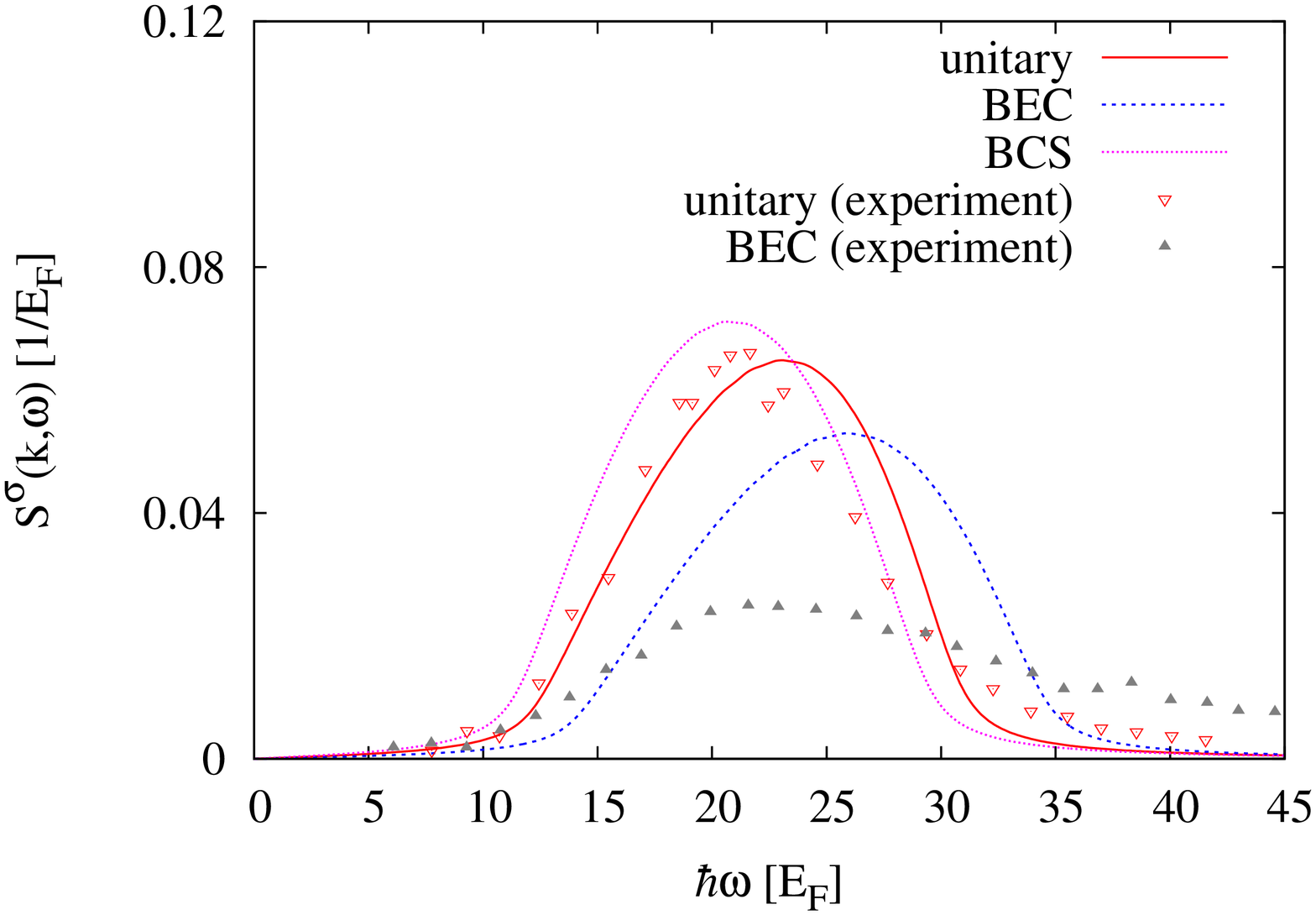}
\end{center}
\caption{(Color online) The dynamic structure function in the density
  (\textit{top}) and spin (\textit{bottom}) channels at unitarity
  ($1/(k_F a)=0$), BEC ($1/(k_F a)=1$), and BCS ($1/(k_F a) =-1$).
  The lines are results from dynamic many-body theory and points are
  experimental data of Ref. \onlinecite{Hoinka2012}. All data
  corresponds to $q=4.5 k_F$.}
\label{fig3}
\end{figure}

In Fig. \ref{fig1}, we show results of $S^{(\rho)}(k,\omega)$ and
$S^{(\sigma)}(k,\omega)$ for the wave vector
$q\,=\,4.5\,k_{\text{F}}$. We have selected this particular $q$ value
because it is the only momentum measured by Bragg
scattering~\cite{Hoinka2012}. In agreement with the experimental
data, the density channel of the dynamic structure function shows a
clear ``Bose'' peak centered at energy $\hbar^2 k^2/4m$ coming from
the scattering off the $\uparrow$-$\downarrow$ pairs. At higher
energies, contribution of single Fermi atoms appear as a broader
response. The importance of correlations at unitarity is best
demonstrated by comparing the response coming from the full theory
with the non-interacting Lindhard function, also shown in
Fig. \ref{fig1}. Our result for the spin channel at unitarity and the
same $q$ value as in the experiment is also shown in Fig. 1 (bottom
panel). As mentioned above, dimers do not contribute to the spin channel
and only atomic contributions appear.  As in the density channel, the
use of many-body theory is crucial to describe the experimental data,
note that the response of a non-interacting system has a completely
different structure and differs substantially from what was observed
in the experiment.

The full momentum-energy dependence of the density and spin response
functions is shown as density plots in Fig.~\ref{fig2}. At low
momenta, $k \alt 4 k_F$, the dimer peak lies inside the particle-hole
($p-h$) band and has a large overlap with the single-atom scattering
contribution. At higher momenta, the ``Bose'' peak is progressively
decoupled of Fermi excitations with a position given by half the
recoil energy. The strength of the spin response is localized inside
the $p-h$ band but interactions make the peak slightly asymmetric and
shifted to higher energy compared with the non-interacting response;
this effect is washed out at momenta $k \agt 4 k_F$.

Bragg scattering~\cite{Hoinka2012,Hoinka2013} was also applied to a
point in the phase diagram located in the BEC side, in particular at
$1/(k_F a)=1$, whereas no experimental data exists in the BCS side. We
have applied dynamic many-body theory to the experimental BEC point
and to the symmetric one in the BCS side, with $1/(k_F a)=-1$, and at
a momentum transfer $q= 4.5 k_F$. The response in the BCS regime is
rather close to that of the free Fermi gas, with only an incipient
Bose peak from the dimer contribution due to its very low
concentration ($< 10$ \%). On the contrary, the dimer contribution
dominates the density channel in the BEC side since the concentration
of dimers is $\sim 90$ \%. Our theory, which produces results in good
agreement with experiment in the unitary limit, seems to worsen in
the BEC regime. This is seen most clearly in the spin response at
$1/(k_F a)=1$ where our result does not show the broad response obtained
from Bragg scattering.

To conclude, motivated by the first measurements of the dynamic
structure function of a two-component resonantly interacting Fermi
mixture, we have developed a theory able to predict its dynamic response.
To this end, dynamic many-body theory is used together with a FN-DMC
input for the static structure factor.  From a direct comparison with
experimental data~\cite{Hoinka2012}, we show that the ideal Fermi gas
model is not sufficient to reproduce correctly the dynamic response
functions. We have formulated a description in terms of a mixture of dimers
(bosons) and atoms (fermions).  This simple model permits to capture
correctly the complicated structure of the density response at the
unitary limit. Our theory departs from the experimental
data, mainly in the spin channel, when the fraction of dimers is large. 
Finally, we provide information on the dynamic
response for different (previously not measured) values of the
momentum in the unitary limit and on the BCS (not measured) side.

We acknowledge partial financial support from the DGI (Spain) Grant
No.~FIS2011-25275  and Generalitat de Catalunya Grant No.~2009SGR-1003 
as well as from the Austrian Science Fund FWF under project I-602.
G. E. A. acknowledges support from the Spanish MEC through the Ramon y Cajal
fellowship program.


\begin{thebibliography}{99}

\bibitem{review} S. Giorgini, L. Pitaevskii, and S. Stringari,  Rev. Mod. Phys.
\textbf{80}, 1215 (2008).

\bibitem{zwierlein} M. W. Zwierlein, A. Schirotzek, C. H. Schunck, and W.
Ketterle, Science \textbf{311}, 492 (2006).

\bibitem{partridge} G. B. Partridge, W. Li, R. I. Kamar, Y. Liao, and R. G.
Hulet, Science \textbf{311}, 503 (2006).

\bibitem{shin} Y. Shin, C. H. Schunck, A. Schirotzek, and W.  Ketterle,
Nature (London) \textbf{451}, 689 (2008).

\bibitem{navon} N. Navon, S. Nascimb\`ene, F. Chevy and C. Salomon, Science
\textbf{328}, 729 (2010).

\bibitem{astra} G. E. Astrakharchik. J. Boronat, J. Casulleras, and S.
Giorgini, Phys. Rev. Lett. \textbf{93}, 200404 (2004).

\bibitem{carlson}  J. Carlson and S. Reddy, Phys. Rev. Lett. \textbf{95}, 060401
(2005).

\bibitem{gandolfi} M. McNeil Forbes, S. Gandolfi, and A. Gezerlis,
Phys. Rev. Lett. \textbf{106}, 235303 (2011).

\bibitem{drut0} J. E. Drut, T. A. L\"ahde, G. Wlaz\l owski, and P. Magierski
 Phys. Rev. A \textbf{85}, 051601 (2012).

\bibitem{salomon} S. Nascimb\`ene, N. Navon, K. J. Jiang, F. Chevy, and  C.
Salomon, Nature (London) \textbf{463}, 1057 (2010). 

\bibitem{mit} M. J. H. Ku, A. T. Sommer, L. W. Cheuk, and M. W. Zwierlein, 
Science \textbf{335}, 563 (2012). 

\bibitem{pilati} S. Pilati and S. Giorgini,  Phys. Rev. Lett. \textbf{100},
030401 (2008).

\bibitem{grigori} G. E. Astrakharchik, S. Giorgini, and J. Boronat, Phys.
Rev. B \textbf{86}, 174518 (2012).

\bibitem{sanner} C. Sanner, E. J. Su, A. Keshet, W. Huang, J. Gillen, R. Gommers, 
and W. Ketterle, Phys. Rev. Lett. \textbf{106}, 010402 (2011). 

\bibitem{sommer} A. Sommer, M. Ku, G. Roati, and  M. W. Zwierlein, Nature
(London) \textbf{472}, 201 (2011).

\bibitem{drut} G. Wlaz\l owski, P. Magierski, J. E. Drut,  A. Bulgac,  
and K. J. Roche, Phys. Rev. Lett. \textbf{110}, 090401 (2013).

\bibitem{pieri} F. Palestini, P. Pieri, and G. C. Strinati, Phys. Rev.
Lett. \textbf{108}, 080401 (2012).

\bibitem{Hoinka2012}  S. Hoinka, M. Lingham, M. Delehaye, and C. J. Vale, Phys.
Rev. Lett. \textbf{109}, 050403 (2012).

\bibitem{Hoinka2013} S. Hoinka, M. Lingham, K. Fenech, H. Hu, C. J. Vale,
J. E. Drut, and S. Gandolfi, Phys. Rev. Lett. \textbf{110}, 055305 (2013).

\bibitem{combescot} R. Combescot , S. Giorgini, and S. Stringari, Europhys.
Lett. \textbf{75}, 695 (2006).

\bibitem{hu} H. Hu and X.-J. Liu, Phys. Rev. A \textbf{85}, 023612 (2012).

\bibitem{kro_theory} H. M. B\"ohm, R. Holler, E. Krotscheck, and M.
Panholzer, Phys. Rev. B \textbf{82}, 224505 (2010).

\bibitem{campbell} C. E. Campbell and E. Krotscheck, Phys. Rev. B
\textbf{80}, 174501 (2009).

\bibitem{godfrin} H. Godfrin, M. Meschke, H.-J. Lauter, A. Sultan, H. M.
B\"ohm, E. Krotscheck, and M. Panholzer, Nature (London) \textbf{483}, 576
(2012). 

\bibitem{Thouless1972}  D. J. Thouless, \textit{The Quantum Mechanics of Many-Body
Systems}, 2nd ed. (Academic press, New York, 1972).

\bibitem{KermanKoonin1976}  A. K. Kerman and S. E. Koonin, Annals of
Physics \textbf{100}, 332 (1976).

\bibitem{KramerSaraceno1980} P. Kramer and M. Saraceno, in \textit{Group 
Theoretical Methods
in Physics}, Lecture Notes in Physics, Vol. 135, edited by
K. Wolf (Springer Berlin / Heidelberg, 1980) pp. 112-121.

\bibitem{ChenClarkSandler1982}  J. Chen, J. Clark, and D. Sandler, 
Zeitschrift f\"ur Physik A Atoms and Nuclei \textbf{305}, 223 (1982).

\bibitem{Krotscheck1982}  E. Krotscheck, Phys. Rev. A \textbf{26}, 3536 (1982).

\bibitem{KroFHNC}  E. Krotscheck, J. of Low Temp. Physics \textbf{119},
103 (2000).

\bibitem{astra2} G. E. Astrakharchik. J. Boronat, J. Casulleras, and S.
Giorgini, Phys. Rev. Lett. \textbf{95}, 230405 (2005).

\end{thebibliography}
\end{document}